\documentclass[prodmode,acmtecs]{acmsmall} 

\usepackage[utf8]{inputenc}
\usepackage[T1]{fontenc}

\usepackage{soul}
\usepackage[colorinlistoftodos,textsize=footnotesize]{todonotes}

\usepackage{amsmath, amssymb,url}
\usepackage{graphicx}
\usepackage{balance}
\usepackage{caption}
\usepackage{subfigure}
\usepackage{comment}
\usepackage{listings}
\usepackage{etex}
\usepackage{makecell}

\usepackage{lipsum}

\usepackage{multirow,ctable}

\usepackage{adjustbox}
\usepackage{array}
\newcolumntype{P}[1]{>{\centering\arraybackslash}p{#1}}
\newcolumntype{M}[1]{>{\centering\arraybackslash}m{#1}}

\usepackage{booktabs}
\usepackage{array}

\newcolumntype{L}[1]{>{\raggedright\arraybackslash}p{#1}}

\lstdefinestyle{base}{
  emptylines=0,
  breaklines=true,
  moredelim=**[is][\color{red}]{@}{@},
}

\usepackage[linesnumbered,ruled,vlined]{algorithm2e}{\small}
\usepackage{algorithm2e}

\SetAlFnt{\small}
\SetAlCapFnt{\small}
\SetAlCapNameFnt{\small}
\SetAlCapHSkip{0pt}
\IncMargin{-\parindent}

\newlength\Origarrayrulewidth

\begin{document}

  \markboth{G. Cascavilla et al.}{OSSINT}

\title{ 
\large{OSSINT - Open Source Social Network Intelligence}\\ \vspace{2mm}
\normalsize{An efficient and effective way to uncover ``private'' information in OSN profiles}
}

\author{Giuseppe Cascavilla
\affil{Sapienza Università di Roma, Italy}
Filipe Beato
\affil{ESAT/COSIC -- KU Leuven and iMinds, Belgium}
Andrea Burattin
\affil{University of Innsbruck, Austria}
Mauro Conti
\affil{Università di Padova, Italy}
Luigi Vincenzo Mancini
\affil{Sapienza Università di Roma, Italy}}

\begin{abstract}
Online Social Networks (OSNs), such as Facebook, provide users with tools to share information along with a set of privacy controls preferences to regulate the spread of information. Current privacy controls are efficient to protect content data. However, the complexity of tuning them undermine their efficiency when protecting contextual information (such as the social network structure) that many users believe being kept private. 

In this paper, we demonstrate the extent of the problem of information leakage in Facebook. In particular, we show the possibility of inferring, from the network ``surrounding'' a victim user, some information that the victim set as hidden. 
We developed a system, named OSSINT (\textbf{O}pen \textbf{S}ource \textbf{S}ocial Network \textbf{INT}elligence), on top of our previous tool SocialSpy, that is able to infer hidden information of a victim profile and retrieve private information from public one.
OSSINT retrieves the friendship network of a victim and shows how it is possible to infer additional private information (e.g., user personal preferences and hobbies). Our proposed system OSSINT goes extra mile about the network topology information, i.e., predicting new friendships using the victim's friends of friends network (2-hop of distance from the victim profile), and hence possibly deduce private information of the full Facebook network.
OSSINT correctly improved the previous results of SocialSpy predicting an average of 11 additional friendships with peaks of 20 new friends. Moreover, OSSINT, for the considered victim profiles demonstrated how it is possible to infer real life information such as current city, hometown, university, supposed being private. 


\end{abstract}

\maketitle
\section{Introduction}
\label{intro}

Online Social Networks (OSNs) are popular web applications that allow users to build connections, establish relationships, and exchange information over the Internet. At the same time, OSNs hold treasure troves of information which are ``insufficiently'' protected by default privacy preferences, by generally applying access control rules to content or users. For instance, Facebook \emph{Folders} and the Google Plus  \emph{Circles} allow the definition of different privacy rules per group, whereas Instagram allows personal pages to be defined as private. However, those privacy preferences are by default hard to use and do not correctly reflect the intentions of users  \cite{FBprivacy,Madejski2012}, which may lead to leakage of information to a wider audience than just their friends. The privacy issues on OSNs has been a topic of interest within the research community demonstrated by several studies \cite{Gross:2005:IRP:1102199.1102214,Beato:2011:SYS:2032162.2032174,Tang:2011,Conti2011,PMID:23479631,Cascavilla:2015:RCI:2808797.2809290}.
Even though the offered privacy controls are somehow effective to protect the data shared, they remain ineffective when protecting contextual information (such as the social network structure) that many users believe being an information kept private, instead leading to the possible leakage of private and sensitive information.




In order to analyse the leakage of information in OSNs, such as Facebook, we propose the use of Open Source INTelligence (OSINT) techniques to extract and infer information from publicly available data sources \cite{OSINT,steele2007open}. In particular, we aim at extracting publicly available data from Facebook and infer information that is set by users to private by means of privacy settings rules. To demonstrate the aforementioned issues, we set up two main targets:
\begin{itemize}
	\item \textbf{Q1}: Would it be possible (and if so, to which extent) to reconstruct personal and supposed hidden friends list?
	\item \textbf{Q2}: Is it possible (and if so, to which extent) to infer personal private information like work, education, hometown, current city of a victim user from his social network (friends and friends of friends)?
\end{itemize}



To respond to our research questions, we built a system, named OSSINT (Open Source Social Network Intelligence), on top of our previous developed SocialSpy \cite{SocialSpy} that exploits the Mutual Content Page (MCP) available in Facebook displaying common content among two users \cite{MutualContentPageFB}.
OSSINT receives as input the list of friends, also called \textit{Friends Found} list (the SocialSpy output), retrieved by SocialSpy from a victim user. Through the MCP, OSSINT retrieves the common friends between the owner of the friends list (victim user) and all the IDs from its \textit{Friends Found} list. OSSINT improves previous results of SocialSpy by predicting multiple-hop friendships (link prediction), such as 2-hop connections (friends-of-friends). In addition, by extracting structural and contextual information OSSINT manages to reconstruct the friendship graph of a victim user along with the importance weight of friends. Hence, it is possible to use the surrounding network (2-hop network composed of friends and friends-of-friends) of a victim profile to infer extra personal information that is supposedly hidden by the privacy preferences. 
Finally, we undeline that our system, OSSINT, does not exploit any Facebook system flaw like those in \cite{FBfriendsList} and \cite{FBfriendsPOSITION} to retrieve victim's information. 

\paragraph{Contribution}
The contribution of this work is manifold. First, we demonstrate that Open Source Intelligence (OSINT), applied to OSNs, allows to retrieve a significant amount of information that users consider, set, and believe is kept private to any prying eyes or third parties, and particularly to other users of the OSN.
Second, we are capable to rebuild the friendship graph of a victim user. Hence from the friendship graph we evaluate the weight of each friendship based on the number of shared friends.
Then, we are also able to extend our finding to multiple-hop connections, in particular showing how to rebuild and learn information of users in the 2-hop network of the victim. Finally, using the friendship graph we exhibit the possibility of rebuilding other private attributes from the 2-hop network, such as personal information, and the possibility to extend to retrieve the full Facebook graph information.
On all our testing cases, OSSINT, correctly improved the results of SocialSpy finding a longer set of friends by link prediction applied to the 2-hop users. With an average of 11 new friendships and peaks of 20 new friends found from the 2-hop network, OSSINT demonstrated the feasibility and the correctness of our assumption. Moreover, for all the victim profiles, OSSINT demonstrated how it is possible to infer real life information supposed being private. 

\paragraph{Organization}
The remaining part of the paper is organized as follows. In Section~\ref{relwork} we review the state of the art. In Section~\ref{system_model} we give a formalization of Facebook and of our system OSSINT. In Section~\ref{privateinfo} we give an overview of OSSINT, showing the interaction between \textit{SocialSpy} and OSSINT, our proposed system. 
In Section~\ref{implementation} we give all the technical details of OSSINT, how it works, what are the tools and the techniques used, an example of the output. In Section~\ref{eval} we present our experimental settings and discuss the results. Finally, Section~\ref{conclusion} draws some conclusions, limitations and future works.

\section{Related Work}
\label{relwork}

There are several studies regarding privacy in Online Social Networks (OSNs) in the literature. These works revealed the lack of privacy and security in OSNs and how simple it is, in some cases,  to get private information about users. Unfortunately, many OSNs users are unaware of the security risks which exist in these types of communications.

Recent studies \cite{Acquisti,Boshmaf:2011:SNB:2076732.2076746} showed how many OSNs users expose personal details about themselves, their friends, and their relationships, whether by posting photos or by directly providing information such as a home address and a phone number.
Furthermore, according to \cite{5231850} and to \cite{Boshmaf:2011:SNB:2076732.2076746}, the Facebook users accept friendship requests from people whom they do not know but with whom they simply have friends in common. By accepting these friendship requests, users unknowingly disclose their private information to strangers.
Obviously the leakage of information is not due only to the friendship requests from strangers but also to the difficulties in correctly tune the privacy settings in the different social networks. 
Almost 13 million users said they had never set, or did not know about Facebook’s privacy tools. Furthermore, 36\% of users share all, or almost all, their wall posts with an audience wider than just their friends \cite{FBprivacy}. According to our studies and experiments, we hardly believe that users are completely aware of actual privacy that OSNs provide them. On the other hand, whenever users know that their profiles have some information leakages, they are often too lazy (or inexperienced) to properly modify the privacy options and make the profile private \cite{Wisniewski2016,Madejski2012}.
However, even when the user configures his profile in a proper way, problems may arise with the privacy settings of third-party applications. For example, the current Graphical User Interface (GUI) of Facebook does not help inexperienced users in understanding what kind of permissions are better to give to an application to keep a good level of privacy \cite{10}. Using this Facebook GUI, a user simply authorizes the application to have access to (all) his data. Once the application is authorized, the data from the user could become publicly available.

Several studies over the years tried to study what are the privacy and security risks originated from the use of OSNs. Interesting surveys, articles and journals are available online \cite{Gross:2005:IRP:1102199.1102214,rohtua,citeulike:13050665,fire2013online} and all of them try to explain what are the risks, what are the problems and how to try to get rid from threats. In particular, the survey of Fire et al. \cite{fire2013online} is an interesting study about threats and solutions in OSNs. The survey is divided in \textit{Threats} and \textit{Solutions} part. In the \textit{Threats} paragraph Fire et al. classify the attacks in four categories: \textit{Classic Threats}, \textit{Modern Threats}, \textit{Combination Threats} and \textit{Threats Targeting Children}. Each category explains in details the available attacks. \textit{Classical Threats} contains attacks like ``Malaware'', ``Phishing Attacks'', ``Spammers'', etc. \textit{Modern Threats} contains attacks like ``Clickjacking'', ``De-Anonymization Attacks'', ``Face Recognition'', ``Fake Profiles'', etc. \textit{Combination Threats} is the combination of classic and modern threats in order to create a more sophisticated attack. Lastly \textit{Threats Targeting Children} that contains attacks like ``Online Predators'', ``Risky Behaviors'', ``Cyberbulling''. 
Then we have the \textit{Solutions} paragraph divided in three main categories: \textit{Operator Solutions}, \textit{Commercial Solutions} and \textit{Academic Solutions}.
In the last category we have the solutions proposed from \cite{lipford2008understanding,Privacy_wizards,privacy_protector,C4PS}.\\
It is, however, possible to find in the literature several studies that aim at showing how it is possible to attack OSNs platforms and retrieve supposedly hidden information \cite{Mislove:2010:YYK:1718487.1718519,Mahmud:2014:HLI:2648782.2528548,7568598}. Counter posed there are studies that try to propose defenses to protect OSNs from attacks. Anyway all these studies are mainly academic contributions that we classify as ``Attacks'' and ``Solutions''.

 
An attack example 
is in \cite{SocialSpy}. Burattin et al. demonstrate how it is possible to retrieve the friends list of a victim user in Facebook, using public information. Results have an average of 25\% of hidden friends found and with peaks of 70\%. The work in \cite{SocialSpy} shows that the lack of information of Facebook gives the possibility to an attacker to retrieve information supposedly hidden.
Then there is the solutions category where we have works that try to mitigate the lack of information from OSNs. 
A solution is proposed by Fire et al. in \cite{friendORfoe} with Social Privacy Protector software (SPP). The SPP software consists of two main parts, namely, a
Firefox add-on and a Facebook application. The two parts provide Facebook users with three different layers of protection. The first layer enables Facebook users to easily control their profile privacy. 
The second layer notifies users of the number of applications installed on their profile which may impose a threat to their privacy. The third layer, a Facebook application, analyzes a user’s friends list.

Lastly we have a less studied problem, the \textit{social graph privacy}: preventing data aggregators from reconstructing large portions of the social graph, composed of users and their friendship links. Knowing who a victim's friends are is a valuable information to retrieve other information related to the victim itself. Personal data privacy can be managed by users using the privacy option from OSNs, while information about a user's interests, places or schools in the social graph can be revealed by any of the user's friends. 
The study in \cite{Bonneau:2009:EFE:1578002.1578005} examines the difficulty of computing graph statistics given a random sample of \textit{K} edges from each node, and found that many properties can be approximated. 
Our study shows how it is possible to reconstruct a friendship graph of a victim profile. Using the SocialSpy tool, first we retrieve the friends list of a victim user. Then we re-run SocialSpy on the IDs from the friends list of our victim in order to have a 2-hop friends lists. Once we have the friends list we use OSSINT to build a friendship graph of our victim ID. The system gives us the possibility to know who are the closest friends of our victim, to de-anonymize those users with an high level of privacy that were not retrievable by SocialSpy, and to profile our victim user using the information available from the 1-hop IDs.

\section{System Model}
\label{system_model}

In this section, we formalise our ``system model'', i.e., the environment where our tool could be immersed, in order to retrieve the information. We assume Facebook as a representative implementation of such environment.

Facebook is composed of different entities. All these entities, together, give the possibility to the final user to perform different actions into such environment. The entities we consider are: \textit{pages}, \textit{users}, \textit{groups} and 
\textit{pictures}. 
\emph{Users} are allowed to perform some actions: become ``friend'' of another user, ``like'' a \emph{page} (and revoke the ``like''), ``join'' a \emph{group} (and leave the group), and ``like'' or ``comment'' pictures (and revoke the ``like'' or delete the ``comment'').
Instead, \textit{pages}, \textit{groups} and \emph{pictures} are ``passive'' entities (i.e., they are managed by \emph{users}).
The set of pages a user likes can be interpreted as the \emph{tastes} of that user. Usually \textit{pages} enable public figures (such as companies, organizations, or celebrities) to create a presence on Facebook \cite{defLikeGroup}.
\textit{Groups} on Facebook are ``places'' where people can share and discuss their common interests and express their opinion around common causes, issues or activities to organize \cite{defLikeGroup}. A group is not always public: tuning its privacy rules, it is possible to set it as public (accessible and searchable to all users in Facebook), private (accessible only if invited; searchable to all users in Facebook) or hidden (accessible only if invited; not searchable to anybody in Facebook).
\textit{Pictures} are usually uploaded by users. On Facebook, it is really difficult to take under control the privacy settings of pictures. There are pictures directly uploaded by a user, pictures where users are tagged, the cover photo (that is always public by default) and profile pictures. Moreover, Facebook gives to its users the possibility to make the profile much more detailed filling fields regarding personal hometown rather than the attended university or the current city. 

More formally, the portion of Facebook that we are going to use in the rest of this paper can be formalized as the tuple:
$
	\textit{Facebook} = (\mathbb{U}, \mathbb{P}, \mathbb{G}, \mathbb{I}, \mathbb{C}, \mathbb{S} ).
$
Specifically, in this notation, we have that:
\begin{itemize}
	\item $\mathbb{U}$ is the set of users. A user $u \in \mathbb{U}$ represents a person. Each person can ``like'' a page $p$, join a group $g$, leave comments into a page, request friendships to other users (accept friendship from other users), upload pictures into his own profile pages.
	\item $\mathbb{P}$ is the set of pages. A page $p \in \mathbb{P}$ is something related to the tastes of a user, i.e., what a user might like.
	\item $\mathbb{G}$ = $(G', n)$ is the multiset that represents groups, where $G' \subseteq \mathbb{U}$ and $n: G' \to \mathbb{N}_{\geq 1}$ is the multiplicity function indicating the number of groups with the same set of users (please note that the same set of users may appear several times). $\mathbb{G}$ represents all the groups on Facebook (please note the same set of users may appears several times). A group, from the Facebook point of view, is a place where a user can promote, share and discuss relevant topics.
	\item $\mathbb{I}$ is the set of pictures. Every picture $i \in \mathbb{I}$ can receive one or more ``likes'' and one or more ``comments''  from a user $u \in \mathbb{U}$. Therefore, it is possible to consider a picture as the pair $i = ({U^l_i}, {U^c_i})$. Where ${U^l_i} \subseteq \mathcal{P}(\mathbb{U})$ is the set of users that liked $i$, and ${U^c_i} \subseteq \mathcal{P}(\mathbb{U})$ is the set of users that commented on $i$.
	\item $\mathbb{C}$ is the set of cities. An hometown $h \in \mathbb{C}$ represents the city where a user $u$ was born. Differently, $cc \in \mathbb{C}$ represents the  current city where a user $u$ currently lives. 
	\item $\mathbb{S}$ is the set of schools. The education field $e \in \mathbb{S}$ represents the attended university of the user $u$. Differently $hs \in \mathbb{S}$ represents the high schools attended by the user $u$.
\end{itemize}

Within our model, a user $u$ is defined as the tuple
$
	u = (\textit{Personal}, U, P, G, I, C, S),
$
where: $\textit{Personal}$ is the set of ``personal''  information (such as the name, the family name, the age), $U \subseteq \mathbb{U}$ is the set of friends of $u$; $L_1,\ldots,L_m \in \mathbb{U}$ are corresponding friends list of $u_1,... ,u_m$; 
$P \subseteq \mathbb{P}$ is set of pages $u$ likes; 
$G \subseteq \mathbb{G}$ is the set of groups $u$ belongs to; and $I \subseteq \mathbb{I}$ is the set of personal pictures (pictures that $u$ uploaded into the social network); $C \subseteq \mathbb{C}$ is the set of cities where $u$ was born and where is currently living; $S \subseteq \mathbb{S}$ is the set of schools where the user $u$ attended the university classes and the high schools lectures. 
Given a user $u$ we can extract each component using a projection operator $\pi$. For example, $\pi_C(u) = cc$ the current city of our victim user, rather then $\pi_E(u) = e$ for the attended university of our victim and, lastly, $\pi_H(u) = h $ for the hometown of our victim.
Due to all these interacting entities, and their complex set of privacy settings, we truly believe that there is a real possibility of a leak of information out of Facebook.

\section{OSSINT: Retrieving Private Information}
\label{privateinfo}

\begin{figure}[h!]
	\centering
	\includegraphics[width=5.48in]{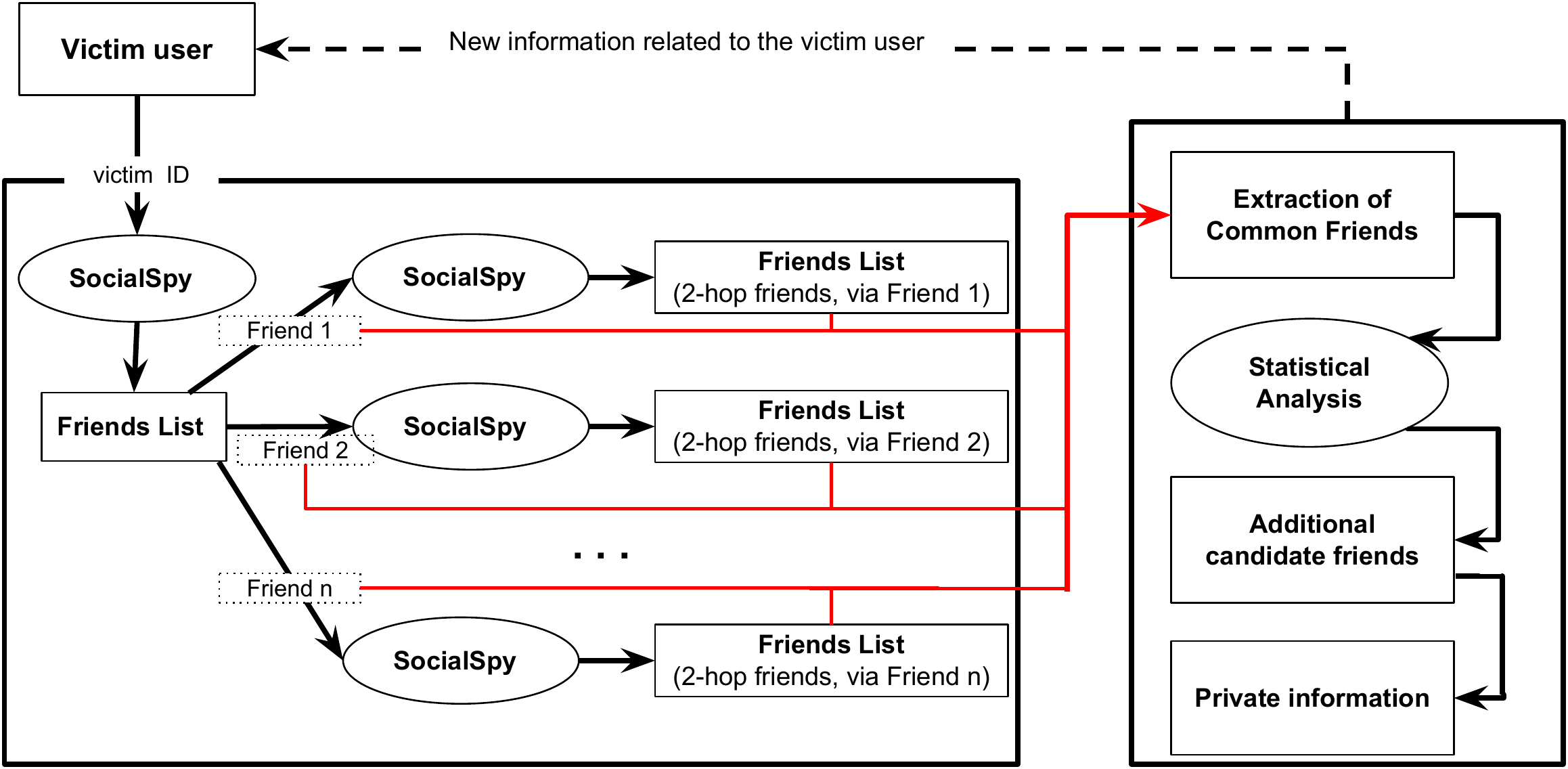}
	\caption {Flow chart and interaction of \textit{SocialSpy} and OSSINT.} \label{fig:ssv2_graph}
\end{figure}

The aim of this work is to show how the friends list can be used as master key to retrieve private information of a victim user that thinks to have set a high level of privacy on a personal Facebook profile. Fig.~\ref{fig:ssv2_graph} gives a graphical representation of our approach and of the involved entities.

The system is composed of two main parts. The first and main part improves the result of our previous tool SocialSpy \cite{SocialSpy}. Then, differently from the previous tool, OSSINT focuses its analysis at 2-hop of distance. To achieve this, we run SocialSpy on the victim user ID and on all the friends returned out of the first run. Therefore, if SocialSpy finds $n$ friends for a victim, we will run it again against all of them, resulting in a total of $n+1$ runs of SocialSpy.

Within a big pool of 2-hop friends (i.e., friends of friends of our victim), OSSINT finds new connection and new friends of our victim user. Moreover exploiting all the found connections, the system is able to retrieve information regarding the current city, the hometown and the education of a victim profile. This constitutes the second part of our system. As mention in our previous work~\cite{SocialSpy}, we decided to have as target the friends list of a victim user because we believed this is one of the most important information on Facebook. With OSSINT we can assert that the friends list of a victim user is, indeed, the weakest link in order to exploit private information. 

In order to retrieve the friends list we based OSSINT on the result of our previous work in~\cite{SocialSpy}. SocialSpy exploits different strategies to fetch this information, however, as we showed in our previous work, the most effective way is to exploit the victim's pictures ($I$, according to our system model). These (or a subset of them), many times, are left publicly available and the strategy we proposed exploits the likes and the comments that each public picture received. In particular, given a picture $i \in I$, belonging to the victim $v$, we can retrieve all the users that liked or commented the picture. Each of them, using Mutual Content Page, is checked for his friendship with $v$. Algorithm~\ref{alg:L_C_pictures} illustrates the steps of SocialSpy in order to retrieve the friends list of a victim user fetching the information from publicly available pictures. 

\begin{algorithm}[h]
	\DontPrintSemicolon
	\KwData{Victim user $v$}
	\KwResult{Set of friends of $v$}
	\BlankLine
	$I \gets \text{set of public images of $v$}$\; 							\label{1st:line:PhotosOf(V)}
	$\textit{CandidateFriends} \gets \emptyset$ \;
	\ForEach{$i \in I$}{
		\tcc{Add candidate friends set all users that liked or commented the image}
		$\textit{CandidateFriends} \gets \textit{CandidateFriends}\ \cup\ U^l_i\ \cup\ U^c_i$ \; 	\label{2nd:line:retrieve_UserLikeCommented}
	}
	\BlankLine
	$\textit{FriendsFound} \gets \emptyset$ \;
	\ForEach {$c \in \textit{CandidateFriends}$								\label{4th:line:foreach_u_like_p}} {
		\tcc{Check friendship with Mutual Content Page}
		\If{$\text{AreFriends}(c, v)$									\label{5th:line:check_u_friend_V}} {
			$\textit{FriendsFound} \gets \textit{FriendsFound}\ \cup\ \{ c \}$ \;			\label{6th:line:FriendsFound=FriendsFound+u}
		}
	}  
	\Return $\textit{FriendsFound}$ \;
	\BlankLine
	\caption{SocialSpy (using ``Strategy~4'' (\textit{S4}), Likes and Comments).} \label{alg:L_C_pictures}
\end{algorithm}
We used the same strategy in OSSINT and we decided to apply it not only at 1-hop of distance from our victim profile, but also at 2-hop (friends-of-friends) of distance. The idea is to improve the pool of IDs on which find new possible friends. Moreover the computational complexity of Algorithm~\ref{alg:L_C_pictures} is $O(n)$ where $n$ is the number of images. 

Algorithm~\ref{alg:potentials} shows how OSSINT extracts 2-hop IDs using the Mutual Content Page (MCP). 


\begin{algorithm}[h]
	\DontPrintSemicolon
	\KwData{Victim user $v$}
	\KwResult{The map of each tuple $(u_i,u_j)$ and its common friends.}
	\BlankLine
	$List\_dicts = [~]$\; 													\label{1st:line:EMPTYpotentialUSERarray}
	\ForEach{$u_i \in socialspy(v)$}{										\label{2nd:line:friendsOFv}
		$dict\_temp = \{~\}$\;		
		\ForEach {$u_j \in socialspy(u_i)$}{								\label{4th:line:friendsOFfriends}
			\tcc{Extracts mutual friends using MCP}
			$dict\_temp[(u_i,u_j)] = M\_F(u_i,u_j)$ \; 						\label{5th:line:mcp}
		}
		$List\_dicts.append(dict\_temp)$\;
	}
	\Return $List\_dicts$\;													\label{7th:line:potentials}
	\BlankLine
	\caption{Set of potentials friends at 2-hop of distance from $v$.} \label{alg:potentials}
\end{algorithm}

Algoritm~\ref{alg:potentials} takes $v$ as input, and for each ID from the friends list of $v$ (line~\ref{2nd:line:friendsOFv}) applies \textit{SocialSpy} (line~\ref{4th:line:friendsOFfriends}) to retrieve the friends list of all the IDs from the friends list of $v$, extracting the common friends. Then, it extracts all the common friends between $u_i$ and $u_j$. $u_i$ belongs to the friends list of $v$ meanwhile $u_j$ belongs to the friends list of $u_i$. Algoritm~\ref{alg:potentials} then, per each $u_j$ from the friends list of $u_i$, extracts the common friends (line~\ref{5th:line:mcp}). The computational complexity of Algorithm~\ref{alg:potentials} is $O(n^m)$ where $n$ is the number of friends of $v$ and $m$ is the maximum number of 2-hop friends. 

\begin{figure}[h]
	\centering
	\includegraphics[width=.99\textwidth]{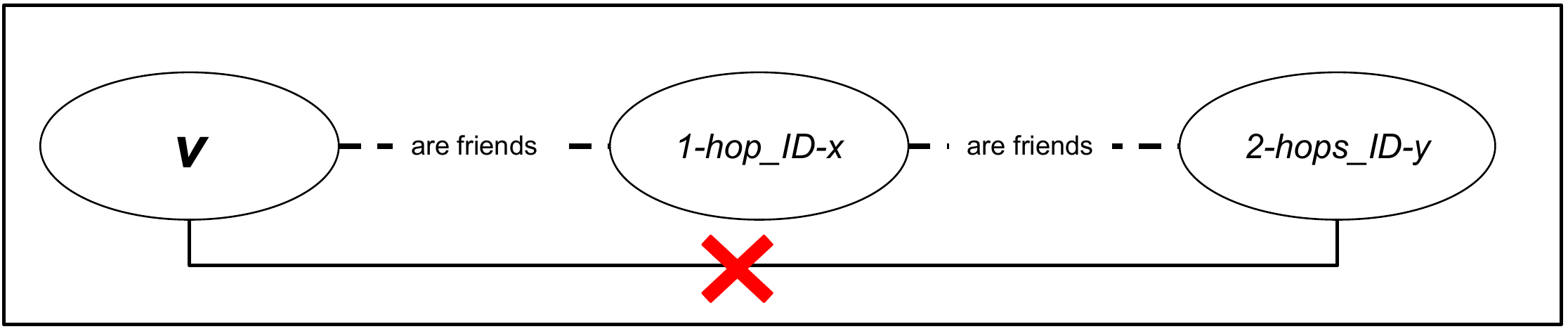}
	\caption{Example of 2-hop friends after the execution of \textit{SocialSpy}. \textit{2-hop\_ID-y} not friend of the victim $v$.}
	\label{fig:not2hopfriend}
\end{figure}

Please note that 2-hop friends are not necessarily friends of our victim user, as reported in the example in Fig.~\ref{fig:not2hopfriend}. 

\begin{figure}[h]
	\centering
	\includegraphics[width=.99\textwidth]{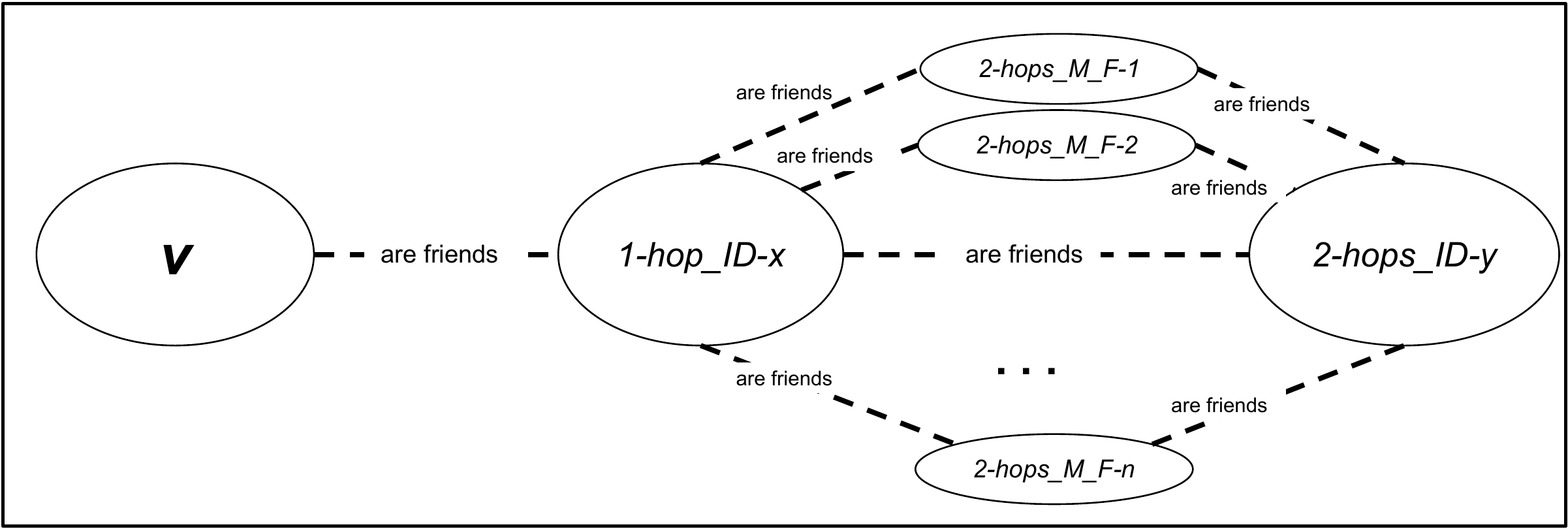}
	\caption{Example of 2-hop candidate friends after the execution of Alg.~\ref{alg:potentials} on $v$.}
	\label{fig:increase2hops}
\end{figure}

Fig.~\ref{fig:not2hopfriend} shows the result after the execution of \textit{SocialSpy} on victim $v$ and on \textit{1-hop\_ID-x}. In order to increase the results of \textit{SocialSpy}, and then the possibilities to find new friendships at 2-hop, we apply Algorithm~\ref{alg:potentials} on \textit{1-hop\_ID-x} and \textit{2-hop\_ID-y}. The result of the execution of Algorithm~\ref{alg:potentials} is depicted Fig.~\ref{fig:increase2hops}.

%

As above mentioned, Fig.~\ref{fig:increase2hops} shows the output after the execution of \textit{SocialSpy} and after the execution of Algorithm~\ref{alg:potentials} on a victim ID. OSSINT produces then a set of friends of $v$, a set of friends for the friends of $v$ and a map data structure where the key of the map is composed by 1-hop friends of $v$ and 2-hop IDs friend of the 1-hop ID. Each key is associated to the list of mutual friends. All the edges from Figures~\ref{fig:not2hopfriend}~and~\ref{fig:increase2hops} are dashed because it is now task of Algorithm~\ref{alg:graphPRODUCTION} to connect all of them. Once all the data are connected, Algorithm~\ref{alg:graphPRODUCTION} will produce the friendship graph. 

\begin{algorithm}[h]
	\DontPrintSemicolon
	\KwData{Victim user $v, List\_dicts$ output of Algorithm~\ref{alg:potentials}}
	\KwResult{$G$ Friendship graph of $v$ at 2-hop}
	\BlankLine
	$V = [v]$\;
	$E = [~]$\; 
	\ForEach {$u_i \in socialspy(v)$}{												\label{1st:line:friendsofV}
			$V.append(u_i)$\;														\label{2nd:line:insertfriendsofV}
			$E.append((v, u_i))$\;													\label{3th:line:connectsfriendsofV}
			\ForEach {$u_j \in socialspy(u_i)$\textbackslash $\{v\}$}{ 				\label{4th:line:friendsoffriends}
				\If {$(u_j \not\in V)$    											\label{5th:line:removeduplicates}}{
				$V.append(u_j)$\;													\label{6th:line:insertF_friendsofV}
				$E.append((u_i, u_j))$\;											\label{7th:line:connectsF_friendsofV}
				\For {$u_k \in M\_F(u_i,u_j)$}{										\label{8th:line:2hopsF_friendsofV}		
					\If {$(u_k \not\in V)$    										\label{9th:line:2hops-ifnotinsertandconnect}}{
					$V.append(u_k)$\;												\label{10th:line:2hops-ifnotinsertandconnect}
					$E.append(u_i,u_k)$\;											\label{11th:line:2hops-ifnotinsertandconnect}
					$E.append(u_k,u_j)$\;											\label{12th:line:2hops-ifnotinsertandconnect}
					}
				}
				}
			}
		}	
	\Return $G = (V,E)$\;															\label{aa}
	\BlankLine
	\caption{Production of the friendship graph $G$ of $v$.}\label{alg:graphPRODUCTION} 
\end{algorithm}

Algoritm~\ref{alg:graphPRODUCTION} is used in order to generate the friendship graph related to our victim user. It takes as input all the outputs generated from the previous algorithms. Starting with the set of friends of $v$, it connects all the IDs (line~\ref{1st:line:friendsofV} to line~\ref{3th:line:connectsfriendsofV}). Once the 1-hop friends are connected it is now the turn to add the 2-hop IDs. Lines~\ref{4th:line:friendsoffriends}, \ref{6th:line:insertF_friendsofV} and \ref{7th:line:connectsF_friendsofV} take care of this step, meanwhile line~\ref{5th:line:removeduplicates} removes the duplicates. Once the graph $G$ results connected at 2-hop, the algorithm uses the common friends in order to insert new edges among the already connected IDs 
(line~\ref{8th:line:2hopsF_friendsofV} to line~\ref{12th:line:2hops-ifnotinsertandconnect}). The output of Algorithm~\ref{alg:graphPRODUCTION} is the friendship graph of the victim $v$ at 2-hop. The computational complexity of Algorithm~\ref{alg:graphPRODUCTION} 
is $O(n^{m*n})$ where $n$ is the number of friends of $v$ and $m$ is the maximum number of 2-hop friends. 

Algorithm~\ref{alg:exractINFO} shows the steps to extract information and compute the rate of each information from each feature (education, hometown and current\_city) retrieved from the friends of $v$.
In order to compute the rate of each element from each feature, Algorithm~\ref{alg:exractINFO} counts how many time the same information appears into the feature taken into account over the number of friends of $v$. Where $\pi_{E}(u_i), \pi_{H}(u_i), \pi_{CC}(u_i)$ are the projections of Education, Hometown and Current\_City.

%

\begin{algorithm}[t]
	\DontPrintSemicolon
	\KwData{Victim user $v$}
	\KwResult{Information and the related rate }
	\BlankLine
	$edu = \{\}$\;
	$hometown = \{\}$\;
	$cur\_city = \{\}$\;
	$M = |socialspy(v)|$\;
	\ForEach{$u_i \in socialspy(v)$}{
	
		\eIf {$(\pi_E(u_i) \not\in edu)$    			\label{1st:line:notinedu}}{					
			$edu[\pi_E(u_i)] = 1/M$\;				\label{2nd:line:compute}}{
			$edu[\pi_E(u_i)] = edu[\pi_E(u_i)] + 1/M$\;				\label{3rd:line:plus1}}
		
		\eIf {$(\pi_H(u_i) \not\in hometown)$    			\label{9th}}{							
			$hometown[\pi_H(u_i)] = 1/M$\;}{
			$hometown[\pi_H(u_i)] = hometown[\pi_H(u_i)] + 1/M$\;}			

		\eIf {$(\pi_{CC}(u_i) \not\in cur\_city)$    			\label{9th}}{							
			$current\_city[\pi_{CC}(u_i)] = 1/M$\;}{
			$current\_city[\pi_{CC}(u_i)] = current\_city[u_i.current\_city] + 1/M$\;}		
	}
	\Return $(edu, hometown, curr\_city)$\;
	\BlankLine
	\caption{Algorithm to extract and rate information from the IDs from the friends list of $v$.}\label{alg:exractINFO}
\end{algorithm}


Algorithm~\ref{alg:exractINFO} computes the percentages of each feature education, hometown, current\_city extracted from the 1-hop users. Line~\ref{1st:line:notinedu} verifies that the information is not already into $edu\{\}$, if not, it add the new feature into $edu\{\}$, otherwise, if the feature is present, Algorithm~\ref{alg:exractINFO} increment the counter for the same feature. The same action is used also for the other features hometown and current\_city.
The computational complexity of Algorithm~\ref{alg:exractINFO} is $O(n)$ where $n$ is the number of friends of $v$.

Algorithm~\ref{alg:scoring} shows the steps to score the likelihood of the 2-hop IDs.
Moreover in Algorithm~\ref{alg:scoring} we use the function showed in Algorithm~\ref{alg:exractINFO}. Indeed this algorithm, first extracts the information regarding education, hometown and current\_city of the 2-hop users, then score the likelihood based on the result given by Algorithm~\ref{alg:exractINFO} applyed on $socialspy(v)$ users. In order to have more reliable likelihood score, Algorithm~\ref{alg:scoring} takes into account also the number of edges (common friends) among $u_j$ (2-hop user) and $v$.

\begin{algorithm}[h]
	\DontPrintSemicolon
	\KwData{Victim user $v$}
	\KwResult{Score of likelihood of 2-hop users associated to $v$}
	\BlankLine
	\tcc{See Algorithm~\ref{alg:exractINFO}}
	$edu, hometown, cur\_city  = proc\_info(v)$\;
	$dict\_scores = \{\}$\;
	$score = 0$\;
	$max = 0$\;
	\ForEach{$u_i \in socialspy(v)$										\label{1st:line:1hop}}{
		\ForEach{$u_j \in (socialspy(u_i)$\textbackslash$socialspy(v))$					\label{2nd:line:2hops}}{
			\If {$u_j \not \in dict\_scores$								\label{3rd:line:noduplicates}}{
				\If {$(\pi_E(u_j) \in edu)$    							\label{4th:line:takevalue}}{
				
					$score = score + edu[\pi_E(u_j)]$\;							\label{5th:line:insertinscore}
				\If {$(\pi_H(u_j) \in hometown)$    						\label{6th:line:takevalue2}}{				
				
					$score = score + hometown[\pi_H(u_j)]$\;						\label{7th:line:xxx}
				\If {$(\pi_{CC}(u_j) \in cur\_city)$    				\label{8th:line:takevalue3}}{				
				
					$score = score + cur\_city[\pi_{CC}(u_j)]$\;			
				$score = (score)/3$\;										\label{10th:line:averagescore}
				$N = |socialspy(u_j) \cap socailspy(v)|$\;					\label{11th:line:numberofedges}
				\If {$max < N$												\label{12th:line:numberofedges}}{
					$max = N$\;												\label{13th:line:numberofedges}
				$dict\_scores[u_j] = (score, N)$\;							\label{14th:line:finalscore}
				}
				}		
				}
				}
			}				
		}
	}
	\ForEach{$u \in dict\_scores$										\label{15th:line:1hop}}{
		$dict\_scores[u] = (dict\_scores[u][0] + (dict\_scores[u][1]/max))/2$\;\label{16th:line:1hop}	
	}
	\Return $dict\_scores$\;
	\BlankLine
	\caption{Algorithm to score the likelihood of IDs at 2-hop.} \label{alg:scoring}
\end{algorithm}


Algorithm~\ref{alg:scoring} takes a 2-hop user (line~\ref{2nd:line:2hops}) and using the function $proc\_info(u)$ extracts the value of the feature. On line~\ref{4th:line:takevalue} the feature retrieved from the 1-hop results is education, on line~\ref{6th:line:takevalue2} the feature is hometown and on line~\ref{8th:line:takevalue3} the feature retrieved is current\_city. The related score from the 1-hop results is added into the variable $score$. Line~\ref{10th:line:averagescore} calculates the average of all the scores previously extracted from the 1-hop users data. Line~\ref{11th:line:numberofedges} extracts the number of edges (or common friends) from $u_j$ (2-hop user) and $v$. Lines~\ref{12th:line:numberofedges},~\ref{13th:line:numberofedges}~and~\ref{14th:line:finalscore} are in charge to find the highest number of shared edges. Based on that all the other values are normalized using the highest one. Lines~\ref{15th:line:1hop}~and~\ref{16th:line:1hop} calculate the final score based on the extracted information and on the number of shared edges.

Once Algorithm~\ref{alg:scoring} returns the $dict\_score$ we compare each score of each ID with a threshold defined looking at the optimum value among our data. If the score of the ID at 2-hop is above our threshold we mark the user ID as ``FRIEND'' if below the ID is marked as ``NOT FRIEND''.
The computational complexity of Algorithm~\ref{alg:scoring} is $O(n^m)$ where $n$ is the number of friends of $v$ and $m$ are the 2-hop friends.


\section{OSSINT: Implementation}
\label{implementation}

%
%


OSSINT is able to fill the gap of SocialSpy between the retrieved friends list and the possibility to have a graphical representation of the whole network of a victim user at 2-hop. The OSSINT system receives as input the list of friends from the IDs found after the execution of SocialSpy on a victim profile. The tool, then, iterates on each ID (from the friends list of $v$) and their personal list of friends in order to extract all the common IDs. After the execution of OSSINT on each ID (from $v$ friends list) and their personal friends list, it connects all the IDs at 1-hop with the IDs at 2-hop. The common friends are then used as edges and also to find new connection among 1-hop and 2-hop IDs.\\ 
In particular the tools used in our system OSSINT are:

\emph{(i)} Mutual Content Page (MCP), i.e., a page that displays which content two users have in common; 

\emph{(ii)} Selenium Web Driver to browse the Facebook MCP pages; 

\emph{(iii)} Graphviz python library to generate a \texttt{.dot} file that represents our output graph; 

\emph{(iv)} Gephi platform for the visualization and the manipulation of the graph.\\

After the iteration of OSSINT on each 1-hop ID and the related friends list, we have multiple output (\texttt{.json}) files where are listed all the common friends among each 1-hop ID and the IDs from the related friends list. OSSINT  can then connect $v$ with its 1-hop friends list and all the 1-hop IDs with the IDs at 2-hop using the common friends found as edges. 



\subsection{Graph Notation}
\label{notation}

Table~\ref{tab:notation} shows the notation used in the next paragraphs. Each color corresponds to a different type of user. The main actors are the victim, the friends of the victim, and the friends of friends of the victim. Moreover, among them, we highlight the common friends, the most relevant, and less relevant 2-hop IDs.
Table~\ref{tab:notation} summarizes and explain all the colours used in the graphs in the following paragraphs. 

\begin{table}[h]
\centering
\caption{Notation of the colours used in the graphs}
\label{tab:notation}
\begin{adjustbox}{width=1\textwidth}
\begin{tabular}{@{}ll@{}}
\toprule
\multicolumn{2}{c}{\textbf{Table of Colours}} \\ \midrule
\includegraphics[width=0.3in]{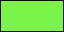}    & Victim user      \\
\includegraphics[width=0.3in]{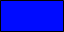}~\includegraphics[width=0.3in]{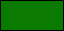}~\includegraphics[width=0.3in]{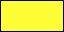}    & Testing IDs and friends of victim user  \\
\includegraphics[width=0.3in]{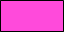}   & Other friends of victim user \\
\includegraphics[width=0.3in]{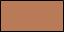}    & Friends of \hspace{1mm} \includegraphics[width=0.3in]{./img/colour4.png} (friends of friends) \\
\includegraphics[width=0.3in]{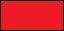}    & Friends of \hspace{1mm} \includegraphics[width=0.3in]{./img/colour1.png} (friends of friends)           \\
\includegraphics[width=0.3in]{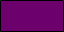}    & Friends of \hspace{1mm} \includegraphics[width=0.3in]{./img/colour6.png} (friends of friends)           \\
\includegraphics[width=0.3in]{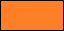}    & Most relevant 2-hop IDs used to predict new friendships of the victim      \\ 
\includegraphics[width=0.3in]{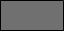}   & 2-hop IDs with only one common friend with the victim (not relevant IDs)\\
\includegraphics[width=0.3in]{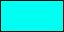}    & 
Common friends between victim user and its testing friends 
\hspace{1mm}~\includegraphics[width=0.3in]{./img/colour4.png}~\includegraphics[width=0.3in]{./img/colour1.png}~\includegraphics[width=0.3in]{./img/colour6.png}
\\
\bottomrule
\end{tabular}
\end{adjustbox}
\end{table}

\subsection{1-Hop Strategy}
\label{strategy}


We are now going to analyze in detail our approach.
The first step of OSSINT is to run SocialSpy on a victim user ID. The SocialSpy tool receives as input a victim ID and returns the friends list of the victim. 
An example of the friends found list returned by SocialSpy is the code in Listing~\ref{code:socialspyfound}.

\begin{lstlisting}[captionpos=b,caption={\textit{SocialSpy} friends found list.}, label={code:socialspyfound},basicstyle=\ttfamily\footnotesize]
Verifying friendship ($user_id_1=`pa***', $user_id_2=`giuseppe.cascavilla')
GET: https://www.facebook.com/pat***?and=giuseppe.cascavilla&sk=friends
*** FRIEND FOUND *** -- 13 An*** (an***)
*** FRIEND FOUND *** -- 99 Va*** (va***)
*** FRIEND FOUND *** -- 272 Al*** (al***)
\end{lstlisting}

\begin{figure}[h!]
  \centering
  \includegraphics[width=2.4in]{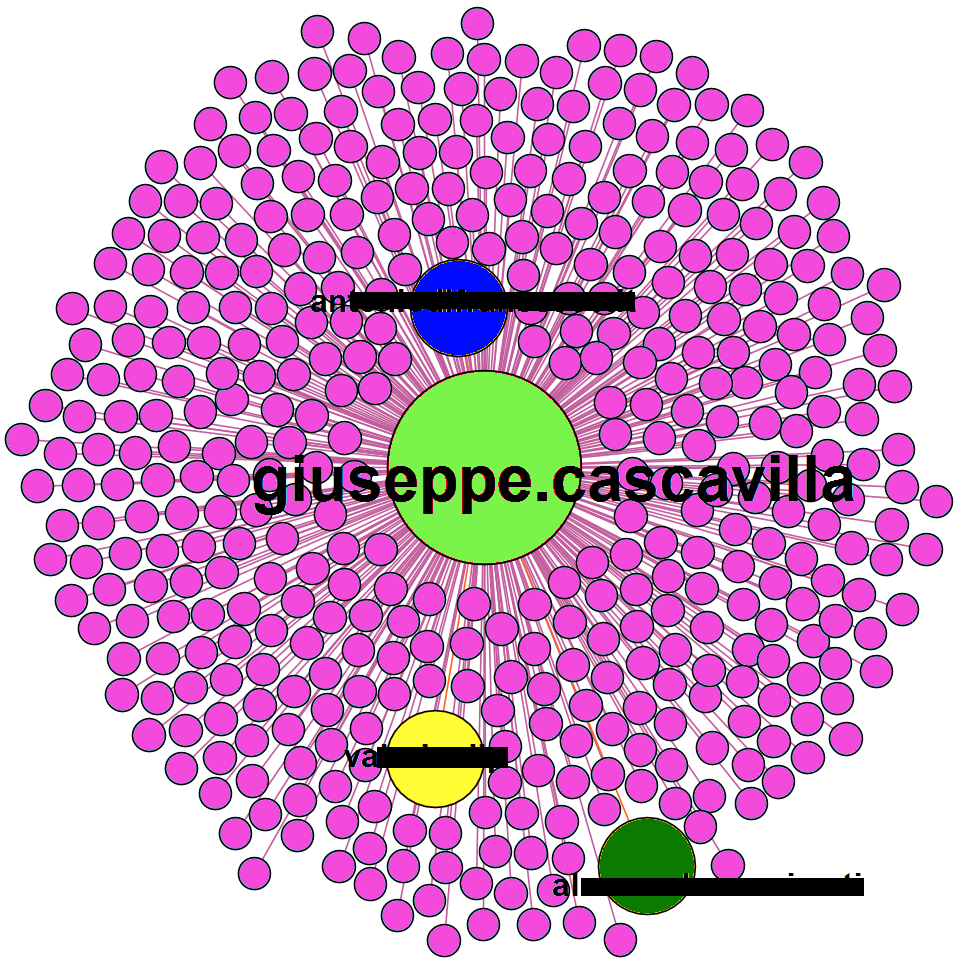}
  \caption {1-hop friendship graph after running \textit{SocialSpy}. \textbf{Green} node: victim ID. 1-hop friend IDs: \textbf{Yellow}, \textbf{Dark Green}, \textbf{Blue} and purple nodes.} \label{fig:friendships_graph}
\end{figure}

Listing~\ref{code:socialspyfound} shows the output of SocialSpy. On the first line SocialSpy checks the friendship using the MCP. The friendship is checked through the written \textit{"Friends since \textbf{[date]}"} available on the MCP. Once the friendship is checked, SocialSpy retrieves the common friends.  
From the retrieved list (Listing~\ref{code:socialspyfound}) we depict the social graph of our victim ID. Indeed, in case of 1-hop (IDs that share the friendship with our victim profile) friends found list, our system OSSINT simply depicts the social graph of our victim. In Fig.~\ref{fig:friendships_graph} an example of graph composed by those friends found at 1-hop. The yellow, dark green and blue nodes are our testing nodes and together with the purple nodes represent the 1-hop friends.

At glance we can notice as the friends found by SocialSpy (Listing~\ref{code:socialspyfound}) are part of our graph depicted in Fig.~\ref{fig:friendships_graph} (dark green, blue and yellow nodes). The green node in the center of our graph is our main victim profile. All the small nodes (purple nodes) are other friends of our victim ID. Yellow, dark green and blue nodes are our testing IDs and friends of our victim.

Once we have the friends list of our victim user we run SocialSpy on each user ID from the retrieved friends found list. The result of this second round is composed by all the friendship graphs of all the IDs from the list of friends of our victim user.

After this second round of SocialSpy we have, per each friend of our victim, a friend found list like the one in Listing~\ref{code:socialspyfound}.

\begin{figure}[h!]
	\subfigure[Friendship graph of victim ID \textbf{va***}. \label{fig:dip}]{\includegraphics[width=.32\textwidth]{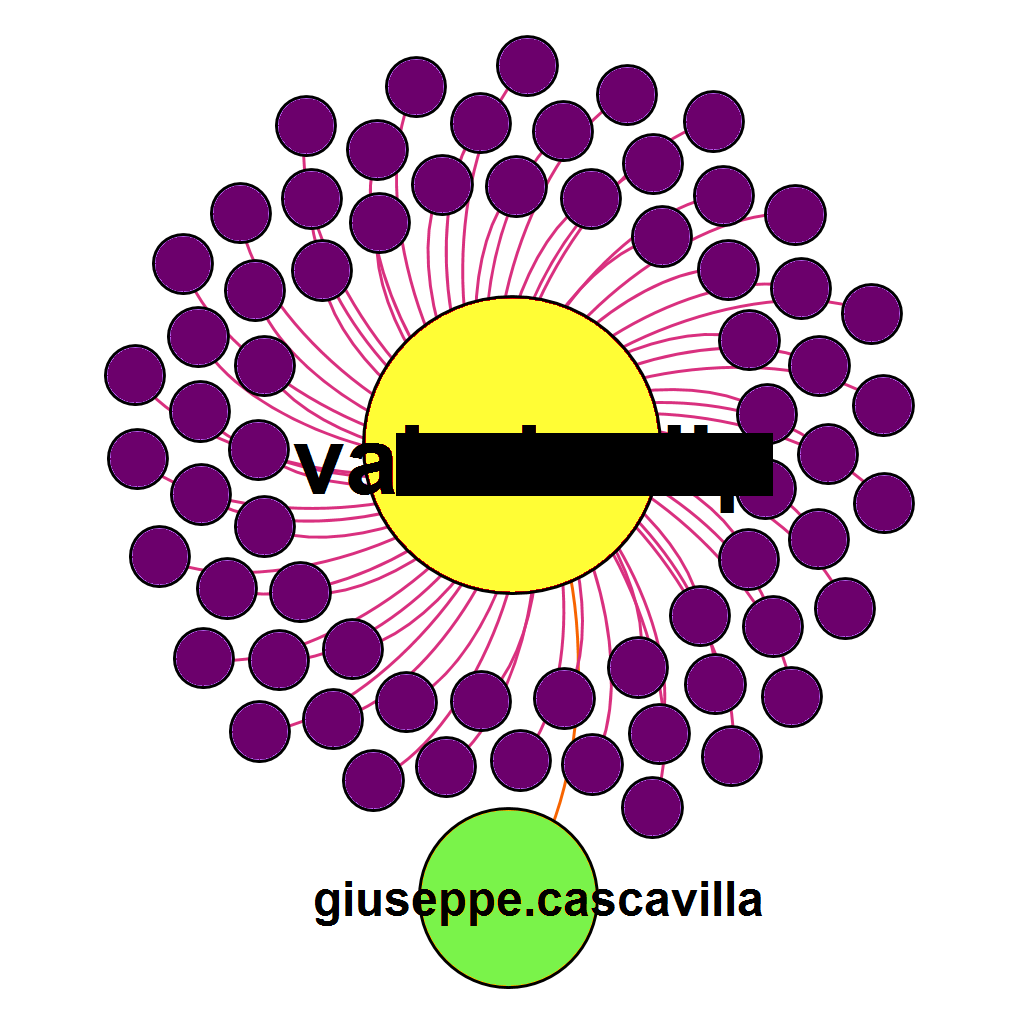}}
	\hfill
	\subfigure[Friendship graph of victim ID \textbf{an***}. \label{fig:difra}]{\includegraphics[width=.32\textwidth]{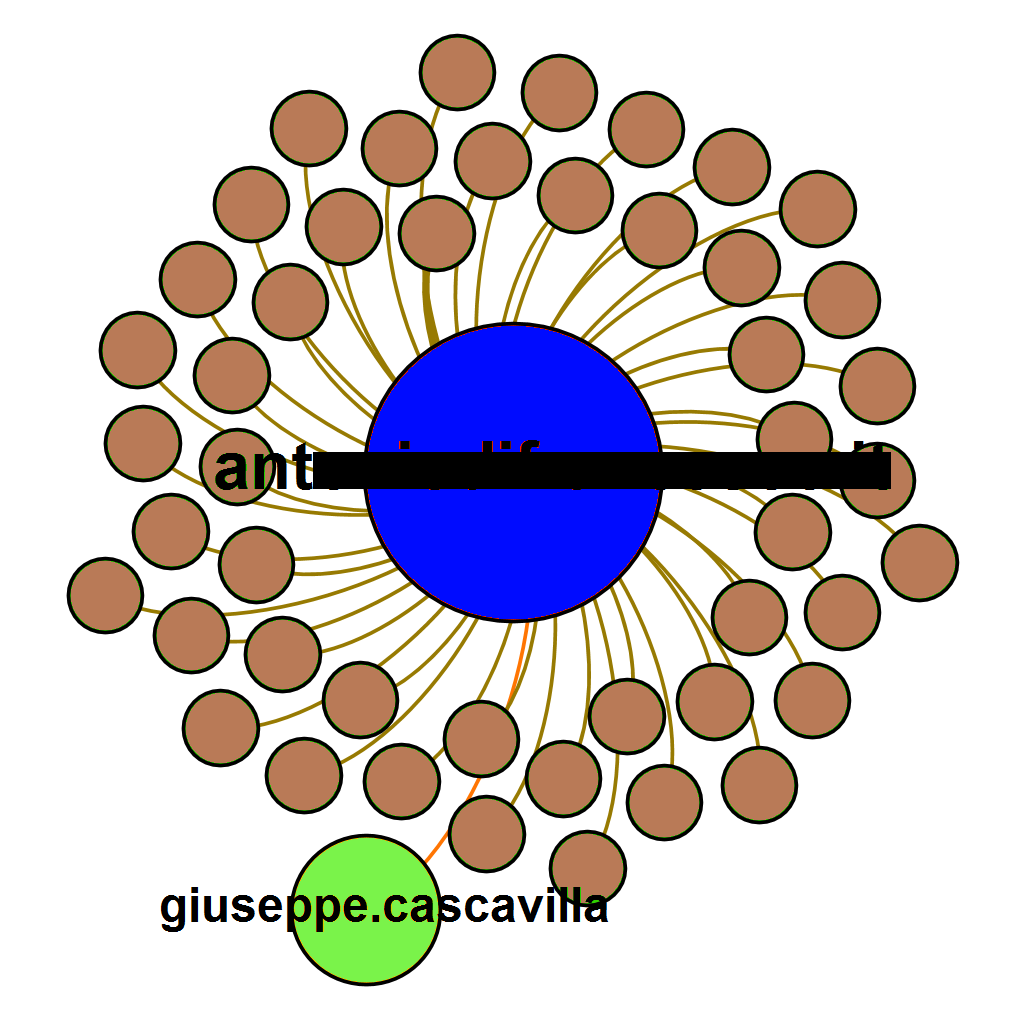}}
	\hfill
	\subfigure[Friendship graph of victim ID \textbf{al***}. \label{fig:cosim}]{\includegraphics[width=.32\textwidth]{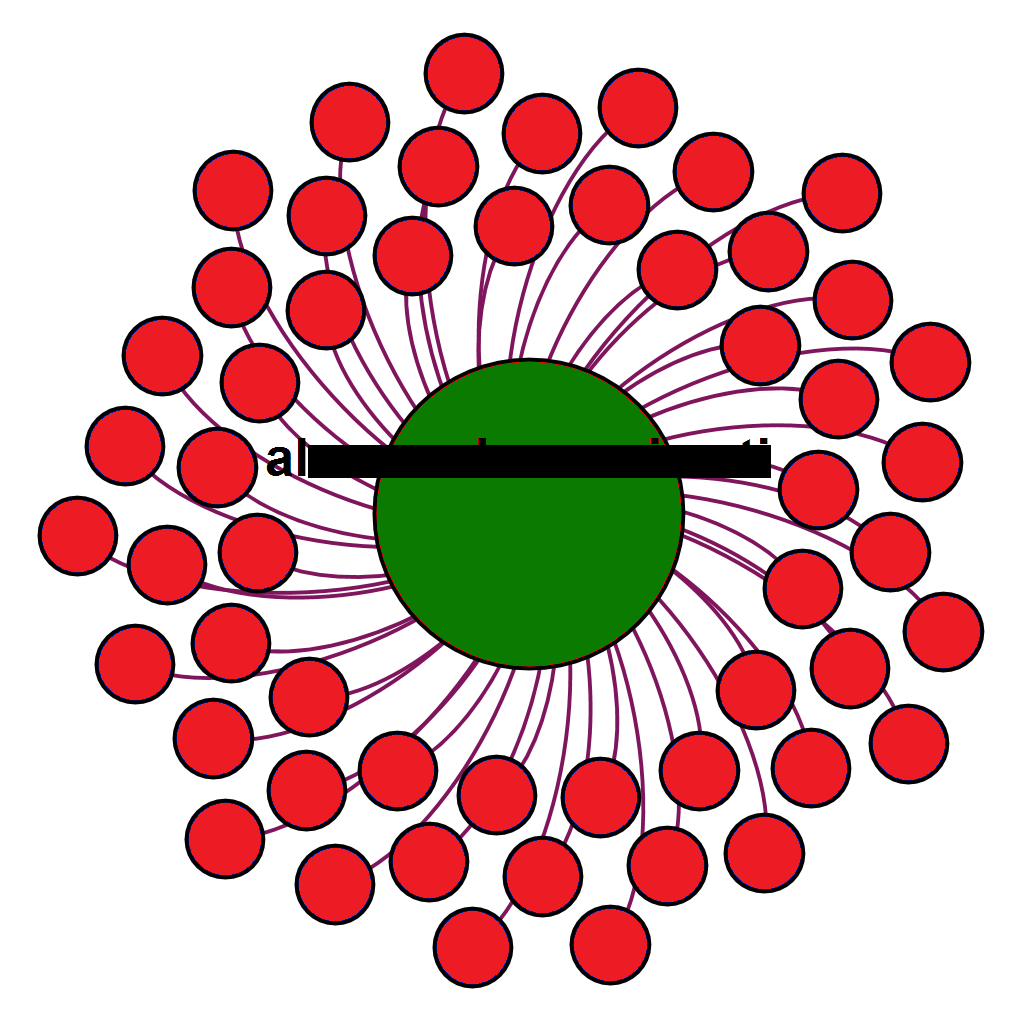}}
	\caption{Graphs of friendship of three IDs that share the friendship with our victim user.}
	\label{fig:friendshipgraphs}
\end{figure}

We can then depict the social graph of our retrieved IDs using OSSINT.
In Fig.~\ref{fig:friendshipgraphs} an example using the testing IDs from Fig.~\ref{fig:friendships_graph}.

The three graphs in Fig.~\ref{fig:friendshipgraphs} depict three different friends network for the three testing IDs from the victim friendship graph (Fig.~\ref{fig:friendships_graph}). The dark green, blue and yellow nodes are the ``new'' victim IDs from the \textit{giuseppe.cascavilla} friends found list. These three victims are friends of our user \textit{giuseppe.cascavilla} and on them we run our SocialSpy tool. The node violet, brown and red represent respectively the friend IDs of our victims (2-hop friends).
Moreover the three graphs in Fig~\ref{fig:friendshipgraphs} give us two different information. 
Graphs in Fig~\ref{fig:dip} and ~\ref{fig:difra} reaffirm the friendship between our victim user \textit{giuseppe.cascavilla} and his friends found IDs. The Fig.~\ref{fig:cosim} shows that our SocialSpy tool, working on \textbf{al***} ID, ``failed'' in finding our victim ID \textit{giuseppe.cascavilla}. However since the friendship in a Facebook social network is an undirected edge, we can assert that exists the friendship between \textit{giuseppe.cascavilla} and \textbf{al***} simply using the  data from \textit{giuseppe.cascavilla}. 

At this stage, our SocialSpy tool retrieved the lists of friends from our victim ID \textit{giuseppe.cascavilla}. SocialSpy, then, retrieved the lists of friends from the 1-hop IDs (Fig.~\ref{fig:friendshipgraphs}) composed of our testing victims \textbf{va***}, \textbf{an***}, \textbf{al***} friends of \textit{giuseppe.cascavilla}. We have then a bunch of friends list files from all our target IDs. At this point our SocialSpy tool becomes limited. Actually, with SocialSpy is not possible to link together all this friends list. Indeed, is not possible to build the friendship graph of the victim  \textit{giuseppe.cascavilla}. Moreover, is not possible to infer private information from the victim ID, \textit{giuseppe.cascavilla}, because of its privacy settings. Lastly, from our studies, we can assert that all these lists of friends can be linked together using shared friendships as edges and in the next paragraphs we will show how we do that.
\subsection{2-Hop Strategy: Proposed System} 
\label{proposed_system}


With the lists of friends retrieved by \textit{SocialSpy}, we are now able to use OSSINT in order to build the friends network of our victim profile. 

The main steps of OSSINT can be summarized as follow:
\begin{enumerate}
    \item extracts all the common friends between each 1-hop ID and the IDs from friends list retrieved by SocialSpy,
    \item produces the text version of the final friendships graph where all the IDs are connected together based on the friendships,
    \item generates a visual representation of the final friendships graph of the victim user at 2-hop of distance
    \item infer information regarding school, hometown and current city of a victim user exploiting the publicly available information of the 1-hop IDs,
    \item produces the list of possible friends of the victim user found at 2-hop of distance (link prediction among victim user and 2-hop IDs) and based on the information retrieved from the 1-hop IDs.
\end{enumerate}

Using the MCP and the Selenium Web Driver, OSSINT is able to retrieve all the mutual links (also known as friendship) among the 1-hop IDs and the friends from their friends list. The output of this iteration are multiple \texttt{.json} files where all the common friends are listed .

An example in Listing~\ref{code:newsystem}.

\begin{lstlisting}[captionpos=b, caption={\texttt{.json} output file after running OSSINT system on \textbf{an***} user ID.}, basicstyle=\ttfamily\footnotesize, label=code:newsystem]
{``https://www.facebook.com/an***?and=ID_1&sk=friends'': [``al***.16'', ``al***'', ``Bo***'', ``caf***'', ``ca***'', ``ce***'', ``cl***'', ``si***'', ``da***'', ``da***'', ``da***'', ``da***'', ``da***''], ``https://www.facebook.com/an***?and=ID_2&sk=friends'': [``as***'', ``mi***'', ``Fo***'', ``rel***'', ``ni***'', ``mi***'', ``lu***'', ``de***'', ``fre***'', ``st***'', ``nu***'', ``de***'', ``di***''], ... }
\end{lstlisting}

OSSINT produces a \texttt{.dot} file right after finishes collecting all the \texttt{.json} files (containing the common friends) among all the 1-hop IDs and the IDs from the 1-hop's friends list. This \texttt{.dot} file is the text version of our final graph. Giving to \textit{Gephi} our generated \texttt{.dot} file we are able to have a visualization of our network at 2-hop of distance.

\begin{figure}[h!]
  \centering
  \includegraphics[width=3.3in]{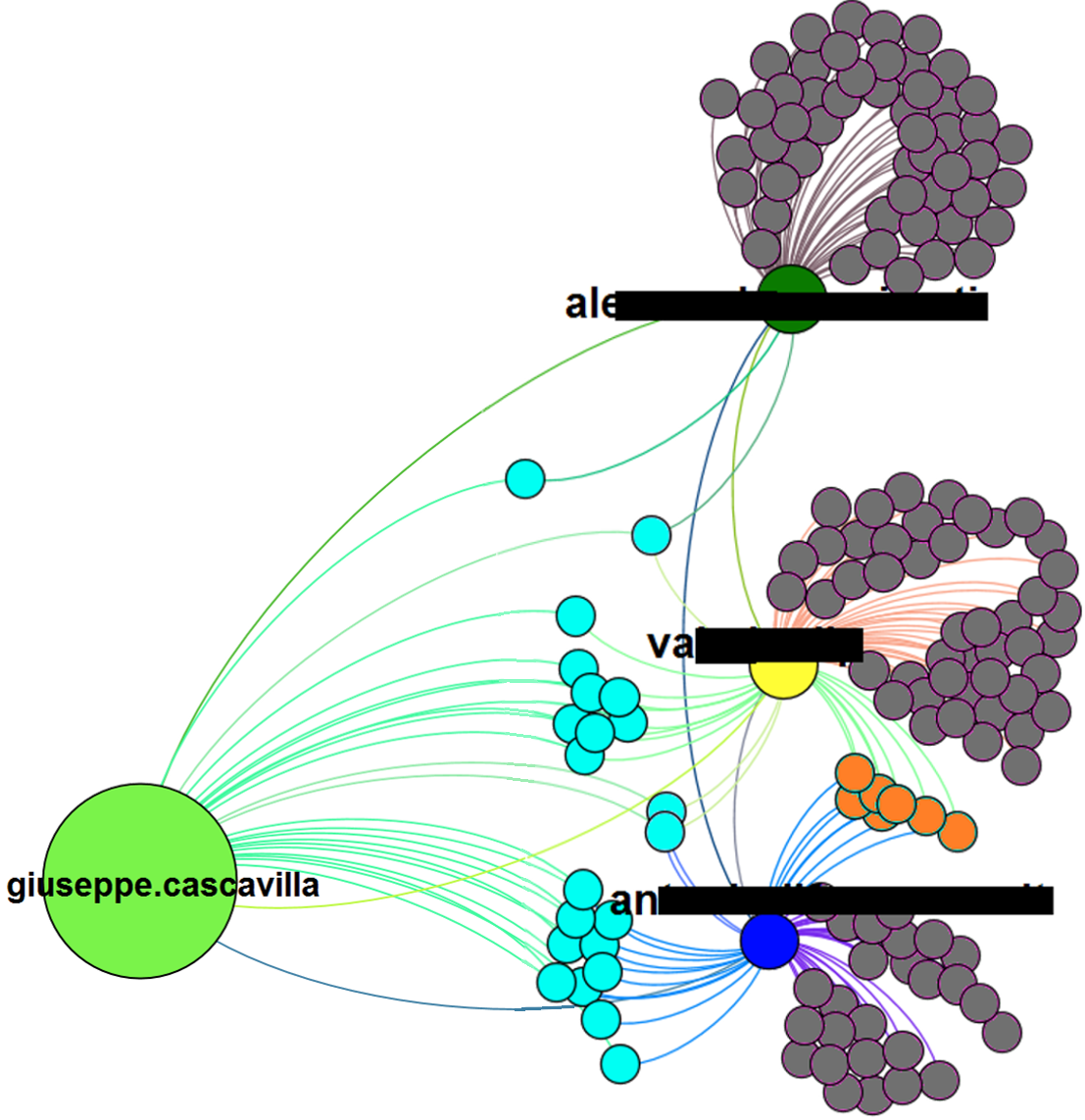}
  \caption {2-hop IDs graph after running \textit{SocialSpy} in conjuction with OSSINT.} \label{fig:friendships_graph_partial}
\end{figure}

In Fig.~\ref{fig:friendships_graph_partial} an example using the data from Figures~\ref{fig:friendships_graph}~and~\ref{fig:friendshipgraphs}.

For readability purpose, from Fig.~\ref{fig:friendships_graph_partial} we removed all the purple nodes, friends of our victim ID \textit{giuseppe.cascavilla} (Fig.~\ref{fig:friendships_graph}). However, it has to be taken into account that, for each purple node, OSSINT makes the same steps as for the dark green, blue and yellow nodes (our testing nodes). However, we now focus only on our three nodes: dark green, blue and yellow and consider them as the only friends of \textit{giuseppe.cascavilla}.
From Fig.~\ref{fig:friendships_graph_partial} we then identify our victim ID (light green node) surrounded by its friends (dark green node, blue node and yellow node). The dark green, blue and yellow nodes are the 1-hop IDs on which we applied SocialSpy to retrieve the 2-hop IDs. The 2-hop IDs are identified by the grey nodes and the orange nodes.
The common friends between the victim ID and its friends are the light blue nodes. From our studies, we can assert that the light blue nodes give us the information regarding who are the closest friends of our victim. Indeed, we strongly believe that more are the common friends among two users, stronger is the friendship among the involved users in a real life. 
We will now focus on the 2-hop IDs to find new friends of \textit{giuseppe.cascavilla} (orange nodes and the grey nodes).

\subsection{2-Hop Strategy: Information Extraction}

The network in Fig.~\ref{fig:friendships_graph_partial} gives us some useful information regarding our victim ID.
\begin{enumerate}
 \item we have the list of friends of our victim ID \textit{giuseppe.cascavilla};
 \item we can infer who are the closest friends of \textit{giuseppe.cascavilla} basing on the number of common IDs between the victim and its friends (higher is the number of common friends, higher is the possibility the two IDs know each other very well in a real life);
 \item we have the list of users at 2-hop of distance;
 \item basing on the number of shared friends between the victim and each of the 2-hop IDs, we can remove the ``useless'' 2-hop users (from Fig.~\ref{fig:friendships_graph_partial} the grey nodes sharing only one edge with only one friend of the victim), keeping the useful ones (from Fig.~\ref{fig:friendships_graph_partial} the orange nodes). 
\end{enumerate}

From the list of friends we retrieved and using the OSINT technique, we extract the information regarding education, hometown and current city, publicly available, from the IDs that share the friendship with \textit{giuseppe.cascavilla} (in our case from dark green, blue and yellow nodes). 
OSSINT produces an \texttt{.xml} file with all the considered information from the friend IDs. In Listing~\ref{code:xmlDATA} an example of the output file.

\begin{lstlisting}[captionpos=b, caption={\texttt{.xml} output file from OSSINT.}, basicstyle=\ttfamily\footnotesize, label=code:xmlDATA]
<?xml version="1.0" encoding="UTF-8"?>
<records> 
  <record>
    <source_url>m.facebook.com/ID/about</source_url>
    <education>School</education>
    <hometown>Hometown</hometown>
    <current_city>Current City</current_city>
  </record>
</records>
\end{lstlisting}

After processing all the data from Listing~\ref{code:xmlDATA}, OSSINT produces an \texttt{.xlsx} file containing one table per each field education, hometown and current\_city. Each table contains all the rates regarding the information from Listing~\ref{code:xmlDATA} retrieved from the 1-hop list of IDs. 
Through this first step we are able to infer the personal information of our victim user. Indeed we truly believe that the personal information of the IDs from the friends list of our victim reflect the private information of our victim ID in a real life. 

\subsection{2-Hop Strategy: Friends Extraction}

After the steps described in Section~\ref{proposed_system}, we now have a huge network, with a lot of IDs at 2-hop of distance and all of them can be possible friends of our victim user. In order to reduce the possibility of errors, because of the huge amount of IDs, we decided to remove from the network, at 2-hop of distance, all the IDs with only one edge. Using as example the graph in Fig.~\ref{fig:friendships_graph_partial}, we remove all the grey nodes that share only one friend with our victim ID \textit{giuseppe.cascavilla}. Differently, we keep all the IDs with more than one common friend with \textit{giuseppe.cascavilla} (Fig.~\ref{fig:friendships_graph_partial} the orange nodes). The decision is based on the fact that the higher is the number of common friends shared between the 2-hop ID and the victim ID, the higher is the probability the victim shares the friendship with the considered 2-hop ID. After removing the one-edge IDs, we apply the OSINT technique on the remaining nodes. 
We extract then the personal information publicly available from the IDs at 2-hop. As for the 1-hop IDs, also in this case OSSINT produces an \texttt{.xml} file with all the personal information regarding our 2-hop IDs. The output file contains the same tags as the one in Listing~\ref{code:xmlDATA}. As before, also in this case the processing of the \texttt{.xml} file produces an \texttt{.xlsx} file containing one table per each field education, hometown and current\_city. However, differently from the previous file, we now use these information to score each user ID. The score is based on how many information fit between the 1-hop IDs statistics and the 2-hop information. In Table~\ref{tab:1hop_data} a scoring example.

%
%
%
\begin{table}[h]
	\centering
	\caption{Data example from 1-hop user IDs}
	\label{tab:1hop_data}
	\begin{adjustbox}{width=0.85\textwidth}
	\begin{tabular}{lrr|lrr|lrr}
		\toprule
		\textbf{Current City} & \textbf{\#} & \textbf{\%} &
		\textbf{Homewtown} & \textbf{\#} & \textbf{\%} &
		\textbf{Education} & \textbf{\#} & \textbf{\%} \\ 
		\midrule
		Padua & 15 & 27\% & Padua & 18 & 13\% & Padua & 39 & 40\% \\[0.5ex]
		Bologna & 5 & 9\% & Rome & 15 & 11\% & Venice & 8 & 10\% \\[0.5ex]
		Paris & 2 & 4\% & Venice & 3 & 3\% &  &  &  \\[0.5ex]
		Madrid & 1 & 2\% &  &  &  &  &  &  \\ 
		\bottomrule
	\end{tabular}
	\end{adjustbox}
\end{table}

Table~\ref{tab:1hop_data} shows the rates regarding the information from the user IDs at 1-hop (friends of our victim user \textit{giuseppe.cascavilla}). These percentages will be part of our scoring mechanism.
In Table~\ref{tab:2hop_data} an example of retrieved data from 2-hop IDs.

\begin{table}[h]
	\centering
	\caption{Data example from 2-hop user IDs}
	\label{tab:2hop_data}
	\begin{adjustbox}{width=0.85\textwidth}
	\begin{tabular}{llllr}
		\toprule
		\textbf{Source ID} & \textbf{Current City} & \textbf{Homewtown} & \textbf{Education} & \textbf{\# Shared Edges} \\ 
		\midrule
		\textit{2-hop\_ID-1} & Padua & Padua & Padua & 8 \\
		\textit{2-hop\_ID-2} & Brussels & Turin & Rome & 3 \\
		2-hop\_ID-3 &  & Venice &  & 10 \\[0.5ex]
		2-hop\_ID-4 & Venice &  & Venice & 2 \\ 
		\bottomrule
	\end{tabular}
	\end{adjustbox}
\end{table}

Table~\ref{tab:2hop_data} shows the information retrieved from the IDs at 2-hop of distance. The score is estimated basing on how many personal information fit with the data from Table~\ref{tab:1hop_data} and how many common friends are shared with our victim user \textit{giuseppe.cascavilla}. All the data are normalized between 0 and 1. If we suppose that the highest number of shared edges is 10, we  normalize the values in column \textbf{\# Shared Edges} dividing by 10 all the values. 
In Table~\ref{tab:scoring_value} a scoring example.

\begin{table}[h]
	\centering
	\caption{Example of scoring based on data from Table~\ref{tab:1hop_data} and Table~\ref{tab:2hop_data}}
	\label{tab:scoring_value}
	\begin{adjustbox}{width=0.70\textwidth}
	\begin{tabular}{lrr}
		\toprule
		\textbf{Source ID} & \textbf{Information Score} & \textbf{Edges Score} \\ 
		\midrule
		\textit{2-hop\_ID-1} & 0.266 & 0.8 \\
		\textit{2-hop\_ID-2} & 0.016 & 0.3 \\
		2-hop\_ID-3 & 0.010 & 1.0 \\
		2-hop\_ID-4 & 0.033 & 0.2 \\
		\bottomrule
	\end{tabular}
	\end{adjustbox}
\end{table}

Table~\ref{tab:scoring_value} shows an example of scoring 2-hop IDs. \textbf{Information Score} is the average score based on the data from Table~\ref{tab:1hop_data}. \textbf{Edges Score} is a score based on the normalization of the value given by the number of shared friends between the 2-hop IDs and the victim user \textit{giuseppe.cascavilla}. 

In order to understand if a 2-hop ID can be highlighted as \textit{``FRIEND''} or \textit{``NOT FRIEND''} of our victim ID \textit{giuseppe.cascavilla}, we compare \textbf{Information Score} and \textbf{Edges Score} values with our \textbf{``Best Value''}. The \textbf{``Best Value''} scores are respectively the optimum value of \textbf{Information Score} and the optimum value of \textbf{Edges Score} from all our victim IDs from all our data. Thus \textbf{``Best Value''} are two fixed parameters, respectively \textbf{Best Information Value} and \textbf{Best Edges Value}, with which to compare the score of our 2-hop IDs. Lastly, if a 2-hop ID is scored with both the values equal or higher than our \textbf{``Best Value''}, the considered ID is highlighted as \textit{``FRIEND''}, otherwise marked as \textit{``NOT FRIEND''}.

\section{Evaluation}
\label{eval}


To evaluate our approach, we conducted several experiments on 8 different real profiles of selected volunteers. We chose only victim IDs that are using the Facebook privacy option. Indeed for our experiments we decided to choose only Facebook profiles with an high level of privacy and with a really low disclosure of information from the personal profile.
This is due to the fact that we truly believe that even with an high accuracy in tuning the privacy options, the disclosure of information can happen through the surrounding network (friends and friend-of-friends).
All the victim IDs are real profiles. 
To each of them we asked to share with us their personal friends list. The personal friends list of our victim profiles has been used as ground truth for our experiments. Thus, to score the precision of our technique, we compared the results of our experiments from each victim ID with the real data from the personal friends list. 

To perform our tests, we logged into Facebook using more than twenty different accounts. All of them are part of a network composed by only those accounts used for the experiments. This is in order to appear to Facebook as real profiles and avoid to be blocked. All the accounts do not share anything with all the victim IDs tested during our experiments. 
As above mentioned, for our experiments we decided to use only users with an high privacy concern. All of them do not show anything about their personal information. The victim profiles have an amount of publicly available pictures between 1 and 13. Most of them are cover pictures. Regarding the friends list, differently from the other information, only 1 profile out of 8 
have a public friends list. However even if the friends list is publicly available, our system OSSINT does not perform any action on it. Indeed, our system OSSINT, re-builds a list of friends using publicly available information and not considering or retrieving data from a publicly available friends list. The friends list then, from the OSSINT point of view, is always considered as private.

\subsection{Experimental Results on Friends Found at 2-Hop}
\label{results_system}
To show the feasibility and effectiveness of our system, we provide a more in depth analysis which demonstrates that with OSSINT we are able to identify new friendships at 2-hop of distance. 
Table~\ref{tab:confusionMatrix} depicts an example of the general confusion matrix we used for our outputs.

\begin{table}[h]
	\centering
	\caption{Confusion matrix}
	\label{tab:confusionMatrix}
	\begin{adjustbox}{width=0.9\textwidth}
	\begin{tabular}{rcc}
		\toprule
		 & \textbf{Not predicted} & \textbf{Predicted} \\ 
		\midrule
		\textbf{Actually not friend} & True Negative (TN) & False Positive (FP) \\
		\textbf{Actually friend} & False Negative (FN) & True Positive (TP) \\
		\bottomrule
	\end{tabular}
	\end{adjustbox}
\end{table}

A confusion matrix \cite{confMATR} contains information about actual and predicted classifications done by a classification system. The entries in the confusion matrix have the following meaning in the context of our study:
\begin{itemize}
\item {\bf TN} is the number of correct predictions that the 2-hop\_ID-\# is not friend of $v$;
\item {\bf FP} is the number of incorrect predictions that the 2-hop\_ID-\# is friend of $v$;
\item {\bf FN} is the number of incorrect predictions that the 2-hop\_ID-\# is not friend of $v$;
\item {\bf TP} is the number of correct predictions that the 2-hop\_ID-\# is friend of $v$.
\end{itemize}

In order to evaluate \textbf{Q1} we run OSSINT system on a set of eight 
users. Differently, to evaluate the quality of OSSINT we used concepts as \textit{precision} and \textit{recall}. The \textit{precision} is the fraction of retrieved instances that are relevant while \textit{recall} is the fraction of relevant instances that are retrieved. \textbf{f1} considers both the precision and the recall of the test to compute the score. 
It can be considered as a weighted average of the precision and recall. 
The value of \textbf{f1} is large when both precision and recall are good, and small when either of them is poor. 
In Formula~\ref{precision},~\ref{recall}~and~\ref{f1} the formulas we used.

\begin{equation}
	\label{precision}
	\textit{Precision =} \frac{TP}{TP+FP}
\end{equation}
\begin{equation}
	\label{recall}
	\textit{Recall} = \frac{TP}{TP+FN} 
\end{equation}
\newcommand\raisepunct[1]{\,\mathpunct{\raisebox{0.5ex}{#1}}}
\begin{equation}
	\label{f1}
	f1 = \frac{2\times \textit{Precision} \times \textit{Recall}}{\textit{Precision} + \textit{Recall}}\raisepunct{.}
\end{equation}


The confusion matrix in Table~\ref{tab:results} summarizes the results of our experiments. 

\begin{table}[h]
	\centering
	\caption{Confusion matrix with average results of the experimentation phase}
	\label{tab:results}
	\begin{adjustbox}{width=0.78\textwidth}
	\begin{tabular}{rcc}
		\toprule
		& \textbf{Not predicted} & \textbf{Predicted} \\
		\midrule
		\textbf{Actually not friend} & 253 & 118 \\
		\textbf{Actually friend} & 28 & 11 \\
		\bottomrule
	\end{tabular}
	\end{adjustbox}
\end{table}

Table~\ref{tab:results} demonstrates how our approach is able to use the information from the 1-hop network in order to retrieve new friendships at 2-hop of distance. 
Our OSSINT system correctly marked 253 IDs as ``NOT FRIEND'' and on the other side, OSSINT has been able to predict an average of 11 new IDs and mark them as ``FRIEND''. From our studies, we can consider the value of ``True Positive'' as a good result that demonstrates how our assumptions are correct. 
Moreover from the above results we can assert that we correctly answered to \textbf{Q1}.

\subsection{Experimental Results on Personal Information}
\label{results_system_personal}

Using the data from Table~\ref{tab:results} we can assert that we are able to find friends at 2-hop of distance. We want now to show the results regarding the personal information like \textit{education}, \textit{hometown}, \textit{current city} of a victim user.
The OSSINT system highlights, indeed, a leak of information using the surrounding network of a victim user. The data showed in Table~\ref{tab:1hop_data} underline the possibility to retrieve personal information of a victim ID that are supposed being private. 
Table~\ref{tab:information_score} shows the percentages of correctness of the user information in the first two positions.

\begin{table}[h]
	\centering
	\caption{Correctness of retrieved information in position Top-One and Top-Two}
	\label{tab:information_score}
	\begin{adjustbox}{width=0.78\textwidth}
	\begin{tabular}{rrrr}
		\toprule
		 & \multicolumn{3}{c}{\textbf{Accuracy}} \\
		\cmidrule{2-4}
		 & \textbf{Current City} & \textbf{Hometown} & \textbf{Education} \\
		\midrule
		\textbf{TOP 1} & 50.00\% & 75.00\% & 75.00\% \\
		\textbf{TOP 2} & 37.50\% & 12.50\% & 12.50\% \\
		\bottomrule
	\end{tabular}
	\end{adjustbox}
\end{table}

 
Through our experiments we can assert that the first and highest two values from the information rate of the 1-hop IDs (Table~\ref{tab:1hop_data}), correspond to the real (from real life) information of our victim ID. From the above results we can assert that we correctly answered to \textbf{Q2} as well. Moreover, we demonstrate the correctness of our assumption showing how it is possible to use the surrounding network of a victim profile in order to retrieve information related to the real life of our victim ID.

\section{Conclusions}
\label{conclusion}

The aim of this work is to present a proof-of-concept approach that demonstrates a significant privacy issue on Facebook. OSSINT is the second building block of our system that started with our previous tool SocialSpy \cite{SocialSpy}. As for the previous work, also here we exploited only tools publicly available in order to reveal information that the victim declared private. OSSINT improves the results of SocialSpy, finding new friendship connections at 2-hop of distance from a victim user. Moreover our study reveals how the list of friends can be exploited in order to retrieve personal information, of a victim profile, like \textit{education}, \textit{hometown}, \textit{current city}. We chose only real victim profiles from the Facebook social network. In order to better stress our system, we decided to choose profiles with a little amount of publicly available information, at least one public picture and not more than thirteen public pictures. This is due to the fact that a smaller amount of public information entail a greater difficulty of retrieving private information of a victim user.  
OSSINT shows the feasibility of the new idea we expressed in \cite{SocialSpy}. The results of our experiments are now able to raise a real concern against Facebook. On the other hand, from our experiments, we hope to create awareness on Facebook users. 

\subsection{Limitations}
As mentioned above, the aim of our project is to present a proof-of-concept and to create awareness on Facebook users. On the other hand we are aware about the small number of victim users of our dataset. However the small amount of victim profiles is due to the fact that Facebook recognizes the pattern of actions. When the same actions are repeated more and more times, Facebook blocks the attacker for some days. In order to avoid to be blocked from Facebook we introduced some tricks like:
\begin{itemize}
\item random delays;
\item fake browsing actions;
\item twenty different fake profiles.  
\end{itemize}

Random delays are used between one action and another in order to appear not as a crawler. Moreover the delays will make our actions as actions from a human profile. The fake browsing actions are used to vary the actions pattern. We, indeed, introduced actions like browsing random pages on Facebook. This contributes to appear as a human that is surfing on Facebook. Moreover, we used twenty different fake profiles in order to split the payload of our experiments. Furthermore, we used the fake profiles to appear to Facebook as different users doing different actions into the social network. Lastly, the exiguous number of victim profiles is due to the stricter rules we applied to select them. In order to be selected, a profile, need to have a really small amount of public information, no public information regarding home town, city and university and not more than 13 public pictures. We chose to apply the aforementioned rules to select our victim profiles in order to better stress our OSSINT tool, and to prove that even on a privacy aware user profiles is possible to retrieve (supposed) private information. 
Moreover, we are aware of our low precision rate however these results come from a long process and OSSINT demonstrated to be able to reduce the research of new friendships into an OSN environment such as Facebook. Lastly, among our 8 victim profiles we have been able to have the results above discussed, however it is to consider that we decided to test the worst case only where our victim users are really privacy aware having few public information available online. The results discussed in section \ref{results_system}, then, can not be used to represent the overall quality of OSSINT. It is, then, not possible to take OSSINT for granted against a random victim profile since that there is a huge variety of profiles in terms of publicly available information and privacy settings.

\subsection{Future Work}
As future work, we want to improve the dataset of victim IDs and rebuild their networks. The experiments, indeed, are still an ongoing task in order to have more data. However, we showed the feasibility of our proof-of-concept and how it is possible to retrieve (supposedly) hidden information from a random victim user. Moreover, we are working on our system, OSSINT, in order to introduce new fake actions (e.g., like and revoke the like of a Facebook page, upload pictures, sharing actions and so on). The aim of introducing new fake operations is to make our OSSINT as close as possible to a real human interactions. This update will let us removing the random delays since that we truly believe that Facebook will not be able to recognize the difference between our system OSSINT and a real human interactions. 
Lastly we will improve the whole system introducing new information to exploit. Indeed we want to use again the friends list of a victim user in order to retrieve new and more information regarding our victim ID. A possible information to exploit can be public pictures where our target ID is tagged rather than comments left on a public friend's wall from our victim or comments where our ID has been tagged.

\section*{Acknowledgment}
Mauro Conti is supported by a Marie Curie Fellowship funded by the European Commission under the agreement n. PCIG11-GA-2012-321980. This work has been partially supported by the TENACE PRIN Project 20103P34XC funded by the Italian MIUR, and by the Project ``Tackling Mobile Malware with Innovative Machine Learning Techniques'' funded by the University of Padua, and by the EU TagItSmart! Project (agreement H2020-ICT30-2015-688061).





\bibliographystyle{ACM-Reference-Format-Journals}
\bibliography{acmFORMAT}








\end{document}